\documentclass[%
 reprint,
superscriptaddress,
 amsmath,amssymb,
 aps,
 pra,
 prb,
 mathtools,physics
]{revtex4-1}
\usepackage{mathrsfs}
\usepackage{graphicx}
\usepackage{dcolumn}
\usepackage{bm}

\begin{document}


\title{A new inflationary Universe scenario with inhomogeneous quantum vacuum}

\author{Yilin Chen}
\email{chenyilin19960823@gmail.com}
\affiliation{College of Physics, Jilin University, 130012, China}
\author{Jin Wang}
\email{jin.wang.1@stonybrook.edu}
\affiliation{College of Physics, Jilin University, 130021, China}
\affiliation{Department of Chemistry and Physics, State University of New York, Stony Brook, NY 11794 USA.}
\affiliation{State Key Laboratory of Electroanalytical Chemistry, Changchun Institute of Applied Chemistry, Changchun, Jilin 130022, China}

\date{\today}

\begin{abstract}
We investigate the quantum vacuum and find the fluctuations can lead to the inhomogeneous quantum vacuum. We find that the vacuum fluctuations can significantly influence the cosmological inhomogeneity, which is different from what previously expected. By introducing the modified Green's function, we reach a new inflationary scenario which can explain why the Universe is still expanding without slowing down. We also calculate the tunneling amplitude of the Universe based on the inhomogeneous vacuum. We find that the inhomogeneity can lead to the penetration of the universe over the potential barrier faster than previously thought.
\end{abstract}

\maketitle

\section{\label{sec:level1}Introduction}
Gravity governs the evolution of the universe. A great breakthrough in gravitational research in the last century is the discovery of general relativity (GR). After Einstein laid down the relationship between the space time geometry through the curvature and the matter within through energy-momentum, the general relativity theory has been applied to many fields, especially in astrophysics and cosmology. Another great achievement of modern physics is quantum mechanics. Based on that, quantum field theory (QFT) emerged, which has been tested in many experiments. Due to the success of relativity in the macroscopic world and quantum mechanics in the microscopic world, it is natural to ask how we can combine them together. In cosmology, this issue becomes more apparent after the birth of the theory of inflationary Universe \cite{guth, Linde}, successful in solving the horizon and flatness problems and eventually quantifying the seeds in terms of the density fluctuations for large scale structure formation and inhomogeneity for the microwave background radiation. In this theory, at the very early history, the Universe expanded exponentially, and the expansion was sustained by the vacuum energy. Despite the success,  one issue remains on how the Universe quits this stage. Many proposals were suggested to resolve this issue\cite{bubble,bubble2}. It turns out rather difficult to complete and have a graceful exit for the old inflationary scenario with multiple bubbles coalescing in a universe suggested by Guth \cite{guth}. Chaotic inflationary scenario has been suggested with essentially one bubble for a universe but forever evolving to avoid the exiting issue.

The idea of the inflationary scenario is to combine the quantum vacuum energy for describing the matter and the Einstein's equation for describing the space time evolution together. Based on the equivalence principle of GR, each form of the energy influences the space-time in the same way. Quantum vacuum brings a new source of energy. Naively,  one can study how the quantum vacuum influence the space-time evolution by simply put the energy momentum tensor for the quantum vacuum on the right hand side and the Einstein space time curvature tensor on the left hand side of the Einstein equation of GR.

Unfortunately, there is an issue once one naively put this two theories of general relativity and quantum mechanics together. This is because there is currently no applicable method for quantizing GR or space-time.  This indicates that quantum mechanics and GR are at totally different footing and are not matched. Therefore, many existing theories suggested certain approximate equations relating these two theories. However, these approaches often show great ambiguity. Firstly, let us look at Einstein's field equation
\begin{align}
    R_{\mu\nu}-\frac{1}{2}Rg_{\mu\nu}=kT_{\mu\nu}.
\end{align}
This equation in the current form would not make sense if $T_{\mu\nu}$ represents the quantum (such as vacuum here) rather than classical matter. This is because the energy-momentum tensor is an operator in quantum world. In quantum mechanics, once we attempt to observe something about a system, the expected observed values corresponds to the average values of the corresponding operator for the observable. Therefore, it seems natural to modify the Einstein equation by changing the energy momentum operator for the average value of it
\begin{align}
    R_{\mu\nu}-\frac{1}{2}Rg_{\mu\nu}=k\langle T_{\mu\nu}\rangle.
\end{align}
This describes one the average level, how the quantum matter influence the space-time evolution. For quantum matter, we know that fluctuations are unavoidable. This is even true for the quantum vacuum. One natural question to ask is how the quantum matter fluctuations influence the space-time evolution. Another way to modify the Einstein equation for taking the fluctuations into account is to take the square of both sides of the Einstein equation and then take the average value on the right hand side
\begin{align}
     \left(R_{\mu\nu}-\frac{1}{2}Rg_{\mu\nu}\right)^2=k\langle {T_{\mu\nu}}^2\rangle.
\end{align}
In fact, (2) and (3) are  equivalent only if the fluctuation of energy-momentum tensor is zero, but this of course is not the case since the energy momentum tensor even for quantum vacuum is not zero.

This example illustrates that before we 'totally' understand quantum gravity, there could be many ways of combining the general relativity and quantum field theory, which are not equivalent at the semi-classical level. Furthermore, these differences in semi-classical treatments are caused by the fluctuations of the quantum field vacuum. The Cosmological Constant Problem \cite{constant} is a good example to demonstrate this issue, where the vacuum energy density predicted by quantum field theory is much larger than the cosmological constant from the observations. A natural question one can ask is whether taking the fluctuation into account can be an effective way to improve the quantitative descriptions of the cosmological evolution driven by the quantum vacuum energy \cite{Unruh}. The idea of considering the quantum vacuum fluctuations was suggested \cite{Unruh} to solve the cosmological constant problem at present. However, quantum fluctuations not only exist in the current Universe with approximately flat space-time but also in the very early history of the Universe. Therefore, it is also important to consider the impacts of quantum vacuum fluctuations on the evolution of the Universe, in particular the early Universe.

In this study, by improving the method developed in reference \cite{Unruh}
, we suggest to go one step further beyond the conventional semi-classical method for combining the GR and the QFT by taking into account of the vacuum fluctuations and based on that reach a new inflationary scenario. In this new scenario, the cosmological constant issue, which arises when trying to combine GR and QFT, can be resolved.

This paper is organized as follows: in section II, we illustrate that the quantum vacuum is not homogeneous, but inhomogeneous caused by quantum fluctuation. In section III, by introducing the modified Green's function, we build up a model to quantify the fluctuations of the quantum vacuum. We study the influence of the quantum vacuum fluctuations and its physical interpretation. In section IV, we consider a simple case and solve the corresponding Einstein's field equation.  In section V, by introducing finite temperature field theory, we take into account of the temperature, which is a key element in cosmological evolution. In section VI, based on our solutions to the cases in section IV and V, we propose a new inflationary scenario, which can help to resolve the cosmological constant problem.  In section VII, we analyze the influence of the inhomogeneous vacuum on the tunneling amplitude of the universe from nothing.

The units and metric signature are set to be $c=\hbar=1$ and $(+,-,-,-)$ throughout. And in this paper, 4-vectors are denoted by light italic type, and 3-vectors are denoted by boldface type.

\section{\label{sec:level1}The quantum fluctuation and inhomogeneous vacuum}

Vacuum energy plays an very important role in the inflationary theory. In this theory, at the very early time, the universe expanded exponentially. In this period, vacuum energy dominated the expansion of the Universe. Usually, the vacuum energy density is treated as a constant, for example, just as in (2), the average value of $T_{00}$ is
\begin{align}
    \langle T_{00}\rangle\sim \frac{\Lambda^4}{16\pi^2},
\end{align}
where $T_{00}=\frac12(\dot{\phi}^2+(\nabla{\phi})^2+m^2\phi^2)$, the energy density of a free scalar field, and the high energy cutoff $\Lambda$ is much greater than the mass in the free scalar field.

By recalling the example in the introduction section, the vacuum fluctuations are not zero. This is because that the vacuum $|0\rangle$ is not the eigen state of $T_{00}$, but  the eigenstate of Hamiltonian $H=\int T_{00}d^3x$ (For the detailed discussions, see Ref.\cite{Unruh}). Therefore, the fluctuations in energy density should be considered.  Due to the vacuum fluctuations, the conventional assumption of homogeneous Universe is only approximately correct. When fluctuations are taken into account, the vacuum is not homogeneous. Thus, a more suitable theory should include the effects of the fluctuations in energy density. To achieve this goal, the strategy we adopt is to modify both sides of the field equation, in order to have the fine structures which are compatible with the fluctuations.

\subsection{\label{sec:level2}Generalizing the FLRW metric}

To describe a homogeneous, isotropic expanding Universe, we introduce Friedmann-Lema\^{i}tre-Robertson-Walker (FLRW) metric\cite{wald},
\begin{align}
    ds^2\!=\!dt^2\!-\!a^2(t)(\frac{dr^2}{1-kr^2}\!+\!r^2d\theta^2\!+\!r^2sin^2 \theta d\varphi^2),
\end{align}
where $k$ can be -1, 0, +1, which indicates the 3-dimensional space is elliptical space (closed), Euclidean space (flat), or hyperbolic space (open) respectively. The factor $a(t)$, known as the scale factor, depends only on $t$. Although the FLRW metric can naturally describe an expanding universe, it cannot describe the inhomogeneous Universe where the inhomogeneity is caused by the vacuum fluctuations. The corresponding resolution is to allow the scale factor $a(t)$ to have spatial dependence.
\begin{align}
    ds^2=dt^2-a^2(t,r)(\frac{dr^2}{1-kr^2}+r^2d\theta^2+r^2sin^2\theta d\varphi^2).
\end{align}
To simplify our model, we only assume that the scale factor has only radius $r$ dependence, so the rotational symmetry is preserved. For the spatial part, $k=1$ is chosen (the reason will be explained in section V). Now, the Ricci tensor for the metric becomes
\begin{align}
    R_{00}&\!=\!-\frac{3\ddot{a}}{a},\nonumber\\
    R_{01}&\!=\!R_{10}\!=\!\frac{2(\dot{a}a'-a\dot{a}')}{a^2},\nonumber\\
    R_{11}&\!=\!\frac{2(1-2r^2)a'}{r(r^2-1)a}+\frac{2a'^2}{a^2}-\frac{2a''}{a}-\frac{2(\dot{a}^2+1)+\ddot{a}a}{r^2-1},\nonumber\\
    R_{22}&\!=\!\frac{R_{33}}{sin^2\theta}\!=\!\frac{r\left[\!(4r^2\!-\!3)a'\!+\!r(r^2\!-\!1)a''\!+\!r\ddot{a}a^2\!+\!2r(\dot{a}^2\!+\!1)a\!\right]}{a},
\end{align}
where the dot represents the derivative with respect to $t$, and the prime represents the derivative with the respect to $r$. Then the Ricci tensor can be substituted into the Einstein's field equation
\begin{align}
    R_{\mu\nu}=8\pi G S_{\mu\nu},
\end{align}
where $S_{\mu\nu}$ is given by the energy-momentum tensor
\begin{align}
    S_{\mu\nu}=T_{\mu\nu}-\frac{1}{2}g_{\mu \nu}T,
\end{align}
where $T$ is the trace ofthe energy momentum tensor $T_{\mu\nu}$.

\subsection{\label{sec:level2}Quantification of the inhomogeneous vacuum}
After transforming the left hand side of equation (8), now we can focus on the right hand side. Following the description of the introduction section, the $S_{\mu\nu}$ which is often taken as the average value (expectation value) of the tensor over the entire space-time should not be taken as a constant due to the vacuum flucutuations. By only taking the expectation values of the energy momentum tensor, some fine structures which come from the fluctuations are lost. Here, our main task is to find a correct $S_{\mu\nu}$ for equation (8), which describes the inhomogeneity of the vacuum fluctuations.
\begin{align}
    \langle0|S_{\mu\nu}|0\rangle=S_{\mu\nu}(\mathbf{x},t).
\end{align}

Before studying how to find the suitable $S_{\mu\nu}$, we introduce the scalar field to describe the matter field in our toy model. For simplicity, $\phi$-4 theory is adopted,
\begin{align}
    \mathscr{L}= \frac{1}{2}g^{\mu\nu}\partial_\mu \phi \partial_\nu \phi-\frac{1}{2}m^2\phi^2-\frac{1}{4!}\lambda\phi^4-\frac{3m^4}{2\lambda},
\end{align}
where $m^2=-\mu^2<0$, Due to the Higgs mechanism, the symmetry is broken spontaneously at low temperatures. The effective potential becomes:
\begin{align}
    V(\phi)=-\frac{\mu^2}{2}\phi^2+\frac{\lambda}{4!}\phi^4+\frac{3\mu^4}{2\lambda}>0.
\end{align}
To be compatible with the modification of the FLRW metric, we reduce a number of degrees of freedom of the field $\phi(t,r)$, and preserve the rotational symmetries just as what we did before.

According to Noether's theorem, the energy momentum tensor for this field is given as
\begin{align}
     T_{\mu\nu}=\partial_\mu \phi \partial_\nu \phi - \frac{g_{\mu\nu}}{2}(\partial^\rho \phi \partial_\rho \phi - \mu^2\phi^2 -\frac{1}{12}\lambda \phi^4-\frac{3\mu^4}{2\lambda}).
\end{align}
Substituting equation (13) in equation (9), we reach
\begin{align}
    S_{00}&=\partial_t\phi\partial_t\phi-V(\phi),\nonumber\\
    S_{11}&=-\frac{1-r^2}{a^2(t,r)}\partial_r\phi\partial_r\phi-V(\phi),\nonumber\\
    S_{22}&=S_{33}=-V(\phi).
\end{align}
Now, we are ready to substitute S into Einstein's equations. At first, let us look at one of the Einstein equations
\begin{align}
    R_{00}=-\frac{3\ddot{a}}{a}=8\pi GS_{00}(r,t)=8\pi G(\langle\dot{\phi}\dot{\phi}\rangle-\langle V(\phi)\rangle).
\end{align}

Next, the main challenge becomes the evaluations of $\langle V(\phi)\rangle$ and $\langle\dot{\phi}\dot{\phi}\rangle$. However, it is impossible to obtain the correct $\langle V(\phi)\rangle$ and $\langle\dot{\phi}\dot{\phi}\rangle$ in conventional methods. The key to obtain the correct expectation values of the potential is to first find out  $\langle0|\phi\phi|0\rangle$. In general, this expectation value is given as
\begin{align}
    \langle0|\phi(x)\phi(x)|0\rangle&=\lim_{x'\rightarrow x}\langle0|\phi(x')\phi(x)|0\rangle\nonumber\\&=\lim_{x'\rightarrow x}G(x',x)
\end{align}
where $G(x',x)$ is the full Green's propagator. To gain insights, the free propagator should be derived at first following the perturbation theory. Due to the spontaneous symmetry breaking, the free propagator cannot be obtained directly. To resolve this issue, we make a shift of the variable, $\phi\rightarrow\phi+\phi_0$, where $\phi_0^2=6\mu^2/\lambda$. Then we have
\begin{align}
    \mathscr{L}=&\frac{1}{2}g^{\mu\nu}\partial_\mu\phi\partial_\nu\phi-\frac{1}{2}(\frac{\lambda\phi_0^2}{2}-\mu^2)\phi^2-\frac{\lambda}{4!}\phi^4-\frac{3\mu^4}{2\lambda}\nonumber\\&+(\mu^2-\frac{\lambda\phi_0^2}{6})\phi_0\phi-\frac{\lambda}{6}\phi_0\phi^3+(\frac{\mu^2}{2}-\frac{\lambda}{4!}\phi_0^2)\phi_0^2
\end{align}
Now it is easy to see that the square of the effective mass of the new field $\phi$ is
\begin{align}
    m_{eff}^2=\frac{\lambda}{2}\phi_0^2-\mu^2=2\mu^2
\end{align}
After shifting the variables, we can derive the solution for $\phi$ and return back to the original variable,
\begin{align}
    \phi=\phi_0+(2\pi)^{-3}\int \frac{d^3k}{\sqrt{2\omega}}[a^{\dagger}_k\psi(x,t)+a_k\psi^*(x,t)].
\end{align}
where $\psi$ satisfies the equation of motion for the free scalar field. For example, in the flat space-time, $\psi(x)=exp(ikx)$, and in this case, the Green's function is
\begin{align}
    G(x',x)&=\phi^2_0+\phi_0\langle\phi\rangle+\phi_0\langle\phi'\rangle+\langle\phi\phi'\rangle\nonumber\\&=\phi_0^2+\int \frac{d^3 k}{(2\pi)^32\omega}e^{ik(x-x')},
\end{align}
where $\omega=\sqrt{k^2+m_{eff}^2}=\sqrt{k^2+4\mu^2}$. In the curved space, the propagator is not as simple as it is in the flat space. Following the Ref.\cite{propagator}, by introducing Riemann normal coordinates and expanding the metric, the propagator in the curved space can also be obtained in momentum space,
\begin{widetext}
\begin{align}
     G(x,x')=&\frac{i\Delta^{1/2}(x,x')}{(4\pi)^{n/2}}\int\frac{d^n k}{(2\pi)^n}e^{ik(x-x')}\left[1+f_1 (x,x')\left(-\frac{\partial}{\partial m^2}\right)+f_2 (x,x')\left(\frac{\partial}{\partial m^2}\right)^2 \right]\frac{1}{k^2+m^2},
\end{align}
\end{widetext}
where $f_1(x,x')$ and $f_2(x,x')$ are certain functions which are related to the curvature tensor. If the curvature is not quite large, the second and third terms in (21) would be much smaller than the first term. In that case, we can establish a perturbative propagator in the curved space-time. For our toy model, we omit $f_1(x,x')$ and $f_2(x,x')$ terms, and just preserve the leading term for the approximate flat space-time (because when the scale factor $a$ is large, the Ricci scalar for FLRW metric is proportion to $a^{-1}$, so the curvature, especially after inflation, is very small.). Meanwhile, the van Vleck determinant $\Delta(x,x')$ should also be one in this case. Thus, the approximate propagator in the curved space-time becomes
\begin{align}
        G(x,x')=\int \frac{d^3k}{(2\pi)^3}\frac{1}{2\omega}e^{-i(\omega\Delta t+\mathbf{k}\cdot\Delta \mathbf{x})},
\end{align}
where the $\Delta t$ and $\Delta x$ are geodesic distance. After that, we can calculate the full propagator. For simplicity, the tree level propagator is considered, and higher order corrections are neglected.

\section{\label{sec:level1}Modifications of the Green's function}

    In this section, we aim at exploring the fluctuations of the vacuum and take this into account in our model by modifying the propagator, then we can establish an effective field theory taking into account of the effects of the fluctuations by introducing the modified Green’s function.

\subsection{\label{sec:level2}Another approach to obtain the propagator}
    Following the previous section, the free propagator for the scalar field is given as
    \begin{equation}
       G(x,x')=\int \frac{d^3 k}{(2\pi)^32\omega}e^{ikx}.
    \end{equation}

   Usually, in quantum field theory, the free propagator can be derived by calculating the Green's function in momentum space and then integrating over $t$ with a suitable contour. However, the two point correlation function obtained by this way is not the expectation value, $\langle0|\phi\phi|0\rangle(x)$, which we expect for the energy-momentum tensor. This is because certain hidden structures inside the correlation functions which can cause the fluctuations are ignored. To show the fine structures of the expectation values for the energy-momentum tensor, here we introduce another approach to obtain the Green's function. We will write down the explicit solution of $\phi$, and then calculate the correlation directly.

   Here, we represent the solution of $\phi$ in terms of the creation and annihilation operators $a$ and $a^{\dagger}$ for the scalar field without interactions
    \begin{equation}
       \phi(\mathbf{x},t)=\int \frac{d^3 k}{(2\pi)^3 \sqrt{2\omega}}(a_k e^{-ikx}+a^{\dagger}_k e^{ikx}).
    \end{equation}
    Then the multiplication of the two $\phi$s becomes
    \begin{widetext}
    \begin{align}
        \phi(\mathbf{x},t)\phi(\mathbf{x'},t')=\int \frac{d^3kd^3k'}{(2\pi)^6}\frac{1}{2\sqrt{\omega\omega'}}[a_k a_{k'} e^{-i(kx+k'x')}+a^\dagger_k a^\dagger_{k'} e^{i(kx+k'x')}
        +a_k a^\dagger_{k'} e^{-i(kx-k'x')}+a^\dagger_k a_{k'} e^{i(kx-k'x')}].
    \end{align}
    \end{widetext}
   Conventionally, in (25), the first two terms do not give contributions to the propagator. This is because that their vacuum expectation values are zero. When we take the expectation value of (25),  $\langle0|\phi(x)\phi(x')|0\rangle$, there would be no difference to the result in (23). However, the crucial thing to consider is how to evaluate the first two terms. To see the significance of the first two terms, here by setting $x'\to x$ and rewriting (25), we have
    \begin{widetext}
        \begin{align}
        \phi^2(\mathbf{x},t)\!=\!\int\! \frac{d^3kd^3k'}{(2\pi)^6}\!\frac{1}{2\sqrt{\omega\omega'}}\big[(a_k a_{k'}\!+\!a^\dagger_k a^\dagger_{k'})\cos(k\!+\!k')x\!+\!i(a^\dagger_k  a^\dagger_{k'}\!-\!a_k a_{k'})\sin(k\!+\!k')x\!+\!(a_k a^\dagger_{k'}\!+\!a^\dagger_k a_{k'})\cos(k\!-\!k')x\!\nonumber\\+\!i(a^\dagger_k a_{k'}\!-\!a_k a^\dagger_{k'})\sin(k\!-\!k')x\big].
    \end{align}
    \end{widetext}
    Similarly, the last two terms indicate zero point energy, and the first two terms are zero when they are taken vacuum expectation value. However, the first two terms are the parts in (26) where the fluctuations emerge
    \begin{align}
        A(\mathbf{x},t)&=\int \frac{d^3kd^3k'}{2(2\pi)^6\sqrt{\omega\omega'}}(a_k a_{k'}+a^\dagger_k a^\dagger_{k'})\cos(k+k')x,\nonumber \\
        B(\mathbf{x},t)&=\int \frac{id^3kd^3k'}{2(2\pi)^6\sqrt{\omega\omega'}}(a^\dagger_k  a^\dagger_{k'}-a_k a_{k'})\sin(k+k')x.
    \end{align}
    Obviously, $\langle A(x,t)\rangle=\langle B(x,t)\rangle=0$. However, $\langle A^2(x,t)\rangle$ and $\langle B^2(x,t)\rangle$ are not zero. In fact, they are the reason why the vacuum state is not the eigen state of the energy density. In this case, $A(x,t)$ and $B(x,t)$ are the key elements to show the 'hidden' structure of $\langle\phi\phi\rangle$. Here,  we have
    \begin{align}
        \langle|A|^2\rangle=&\int \frac{d^3k'd^3k}{(2\pi)^6}\frac{1}{2\omega'\omega}\cos^2(k+k')x,\nonumber\\\langle|B|^2\rangle=&\int \frac{d^3k'd^3k}{(2\pi)^6}\frac{1}{2\omega'\omega}\sin^2(k+k')x.
    \end{align}

   A simplest way to preserve the part which quantifies the fluctuation   is given as
    \begin{equation}
       G(\mathbf{x},t)=\frac{\Lambda^2}{8\pi^2}+\sqrt{\langle |A|^2(\mathbf{x},t)\rangle}+\sqrt{\langle |B|^2(\mathbf{x},t)\rangle}
    \end{equation}
    Here, we  simply preserve the latest non-zero order of $\langle|A(\mathbf{x},t)|^{n}\rangle$ and $\langle|B(x,t)|^{n}\rangle$, so that to leave all the parts in (26) 'survived' after taking the expectation value. Fortunately, $G(\mathbf{x},t)$ is convergent when $x$ and $t$ go to infinity and
    \begin{align}
    \lim_{x,t\rightarrow\infty}G(x,t)\sim\Lambda^2\nonumber\\
    G(0,0)\sim\Lambda^2
    \end{align}
    $G(0,0)$ is related to $\Lambda^2$.  In order to be independent of the high momentum cutoff, we can rewrite this result by subtracting $G(0,0)=G_0$, so
    \begin{align}
       G_R(\mathbf{x},t)=G(\mathbf{x},t)-G_0\nonumber=&\sqrt{\langle A^2(\mathbf{x},t)\rangle}+\sqrt{\langle B^2(\mathbf{x},t)\rangle}\nonumber\\&-\sqrt{\langle A^2(0,0)\rangle}-\sqrt{\langle B^2(0,0)\rangle}
    \end{align}
    Now, the new Green's function is totally composed of the terms of the fluctuations $A(\mathbf{x},t)$ and $B(\mathbf{x},t)$ and start at zero. Meanwhile, the large constant which is proportion to $\Lambda^2$ is eliminated.  This constant is the vacuum energy which is non-observable, due to the QFT.

    \subsection{\label{sec:level2}The interpretation of modified Green's function}
    To gain the modified Green's function, equation (28) should be computed. However, the integration is not simple, Here is a way which can be used to estimate the integrals for $\langle|A(\mathbf{x},t)|^2\rangle$ and $\langle|B(\mathbf{x},t)|^2\rangle$. At the first, divide the interval of the integration
    \begin{align}
        \langle |A|^2(\mathbf{x},t)\rangle&=(\int^{\Lambda}_{\Lambda_0}+\int^{\Lambda_0}_0) \frac{d^3kd^3k'}{(2\pi)^6}\frac{1}{2\omega\omega'}\cos^2(k+k')x,\nonumber \\
        \langle |B|^2(\mathbf{x},t)\rangle&=(\int^{\Lambda}_{\Lambda_0}+\int^{\Lambda_0}_0) \frac{d^3kd^3k'}{(2\pi)^6}\frac{1}{2\omega\omega'}\sin^2(k+k')x.
    \end{align}
    For the first part of integrals, because $\Lambda>>2\mu$, so we have
    \begin{align}
        \omega=\sqrt{2\mu^2+k^2}\approx k.
    \end{align}
    In this case, the integrals become
    \begin{widetext}
    \begin{align}
        \langle A^2_1\rangle=\int^{\Lambda}_{\Lambda_0} \frac{d^3kd^3k'}{(2\pi)^6}\frac{1}{2\omega\omega'}\cos^2(k+k')x\approx \int^{2\pi}_0 d\varphi d\varphi'\int^{\pi}_0 \sin\theta \sin\theta'd\theta d\theta'\int^{\Lambda}_{\Lambda_0} \frac{kk'dkdk'}{2(2\pi)^6}\cos^2(k+k')x,\nonumber\\
        \langle B^2_1\rangle=\int^{\Lambda}_{\Lambda_0} \frac{d^3kd^3k'}{(2\pi)^6}\frac{1}{2\omega\omega'}\sin^2(k+k')x\approx \int^{2\pi}_0 d\varphi d\varphi'\int^{\pi}_0 \sin\theta \sin\theta'd\theta d\theta'\int^{\Lambda}_{\Lambda_0} \frac{kk'dkdk'}{2(2\pi)^6}\sin^2(k+k')x.
    \end{align}
    \end{widetext}
    To see the result of integrals in (34) more clearly, setting $\mathbf{x}=0$ and keeping $\langle A^2_1\rangle$ and $\langle B^2_1\rangle$ as functions  only with variable $t$. Here we have
    \begin{widetext}
    \begin{align}
        \langle A_1^2(0,t)\rangle=
        \frac{1}{32 t^4}\Big[4t^4\left(\Lambda^2-\Lambda_0^2\right)^2+\left(1-4\Lambda_0^2t^2\right) \cos (4 \Lambda_0  t)+\left(1-4\Lambda^2t^2\right) \cos (4 \Lambda t)+2\left(4 \Lambda_0  \Lambda  t^2-1 \right)\cos (2 t ( \Lambda_0  +\Lambda ))\nonumber\\ \qquad\qquad -4\left( \Lambda_0+\Lambda \right) t \sin (2 t ( \Lambda_0 +\Lambda ))+4  \Lambda_0 t \sin (4  \Lambda_0   t)+4 \Lambda  t \sin (4 \Lambda  t)\Big],\nonumber\\
        \langle B_1^2(0,t)\rangle=\frac{1}{32t^4}\Big[4t^4\left(\Lambda^2-\Lambda_0^2\right)^2+\left(4 \Lambda_0^2 t^2-1\right) \cos (4 \Lambda_0  t)+\left(4 \Lambda ^2 t^2-1\right) \cos (4 \Lambda  t)+2 \left(1-4 \Lambda_0  \Lambda  t^2\right) \cos (2 t (\Lambda_0 +\Lambda ))\nonumber\\\qquad\qquad +4 \left(\Lambda_0+\Lambda\right)  t \sin (2 t (\Lambda_0 +\Lambda ))-4 \Lambda_0  t \sin (4 \Lambda_0  t)- 4t \Lambda\sin (4 \Lambda  t)\Big].
    \end{align}
    \end{widetext}

    After direct calculations, $\langle A^2_1\rangle$ and $\langle B^2_1\rangle$ have limits where
    \begin{align}
        \lim_{t\rightarrow\infty}\langle A_1^2\rangle=\lim_{t\rightarrow\infty}\langle A_2^2\rangle=\frac{\left(\Lambda^2-\Lambda_0^2\right)^2}{64\pi^4},
    \end{align}
    At the origin, we also have
    \begin{align}
        \langle A^2_1(0,0)\rangle&=\frac{(\Lambda^2-\Lambda_0^2)^2}{32\pi^4},\nonumber\\
        \langle B^2_1(0,0)\rangle&=0.
    \end{align}

    Then, think about the next part of integrals. In the next part of integrals, because $\Lambda_0<<\mu$, so
    \begin{align}
        \omega=\sqrt{2\mu^2+k^2}\approx \sqrt{2}\mu
    \end{align}
    Similarly, we have
    \begin{widetext}
    \begin{align}
        \langle A^2_2\rangle=\int^{\Lambda_0}_0 \frac{d^3kd^3k'}{(2\pi)^6}\frac{1}{2\omega\omega'}\cos^2(k+k')x\approx\int^{2\pi}_0 d\varphi d\varphi'\int^{\pi}_0 \sin\theta \sin\theta'd\theta d\theta'\int^{\Lambda_0}_0 \frac{k^2k'^2dkdk'}{4(2\pi)^6\mu^2}\cos^2(k+k')x,\nonumber\\
        \langle B^2_2\rangle=\int^{\Lambda_0}_0 \frac{d^3kd^3k'}{(2\pi)^6}\frac{1}{2\omega\omega'}\sin^2(k+k')x\approx\int^{2\pi}_0 d\varphi d\varphi'\int^{\pi}_0 \sin\theta \sin\theta'd\theta d\theta'\int^{\Lambda_0}_0 \frac{k^2k'^2dkdk'}{4(2\pi)^6\mu^2}\sin^2(k+k')x.
    \end{align}
    \end{widetext}
    For simplicity, setting $x=0$, we can easily evaluate equation (39) directly
    \begin{align}
        \langle A^2_2(0,t)\rangle=\frac{1}{288\pi^4}\frac{{\Lambda_0}^6}{\mu^2}\cos^2(2\sqrt{2}\mu t),\nonumber\\
        \langle B^2_2(0,t)\rangle=\frac{1}{288\pi^4}\frac{{\Lambda_0}^6}{\mu^2}\sin^2(2\sqrt{2}\mu t).
    \end{align}

    Combining (36) and (40), the modified Green's function $G_R(0,t)$ is obtained. Because $\Lambda>>\mu>>\Lambda_0$, the approximate result is given as
    \begin{align}
        G_R(0,t)\approx \sqrt{\langle A^2_1(0,t)\rangle}+\sqrt{\langle B^2_1(0,t)\rangle}-\frac{\Lambda^2}{4\sqrt{2}\pi^2}.
    \end{align}
    In this case, the limit when time goes infinity becomes
    \begin{align}
        \lim_{t\rightarrow\infty}G_R\approx \frac{(2-\sqrt{2})\Lambda^2}{8\pi^2}.
    \end{align}
        \begin{figure}
        \includegraphics[width=0.4\textwidth]{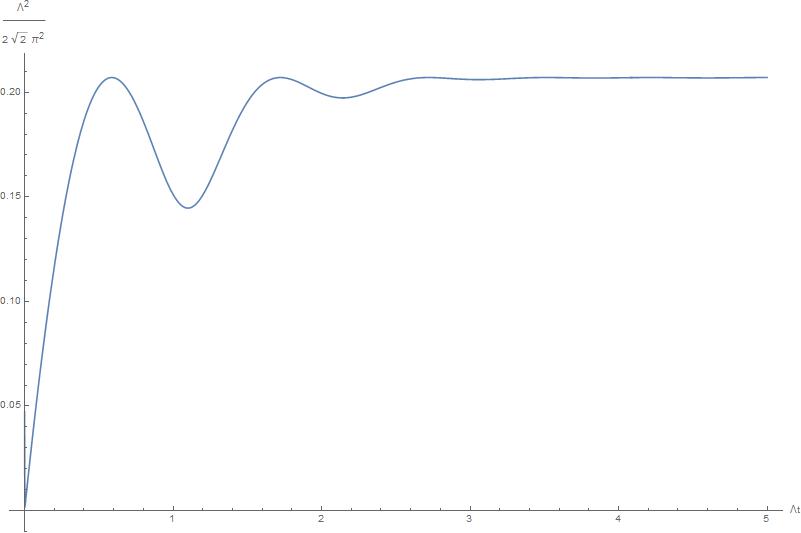}
        \caption{Approximate $G_R(0,t)$ in (41).}
        \label{fig:my_label}
    \end{figure}
    To check the validity of the approximate modified Green's function derived here, we calculate $G_R(0,t)$ numerically.
    \begin{figure}
        \includegraphics[width=0.4\textwidth]{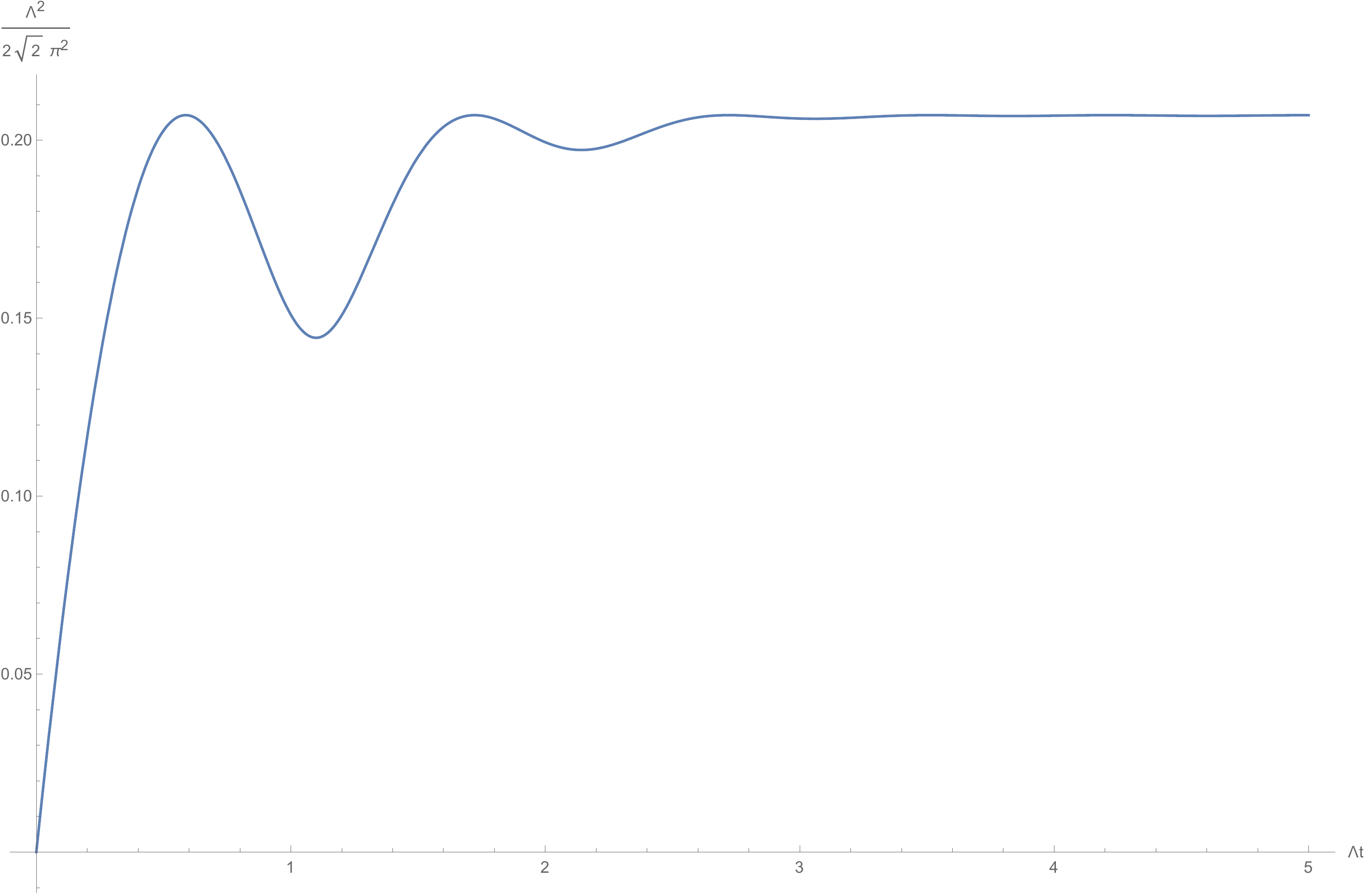}
        \caption{$G_R(0,t)$ derived by numerical calculation.}
        \label{fig:my_label}
    \end{figure}

    In Fig.2, the numerical result is seen to be close to the result in Fig.1 which is plotted with the approximate Green's function. The amplitude of the oscillation of $G_R(0,t)$ remains nearly a constant in Fig.1 at initial times, however, when the time becomes very long, the amplitude of the oscillations of our approximate result approaches zero. The maximal values of the two results derived by different methods are slightly different. In fact, this is not hard to explain. In (41), we drop the $\langle A^2_2(0,t)\rangle$ and $\langle B^2_2(0,t)\rangle$ terms, which influence the amplitude and the value of peak. Roughly speaking, comparing to the value of the Green's function's limit when $t\rightarrow\infty$, the oscillation's amplitude is much smaller comparing to the overall value. Therefore, it can be omitted.
    \begin{figure}[t]
        \includegraphics[width=0.4\textwidth]{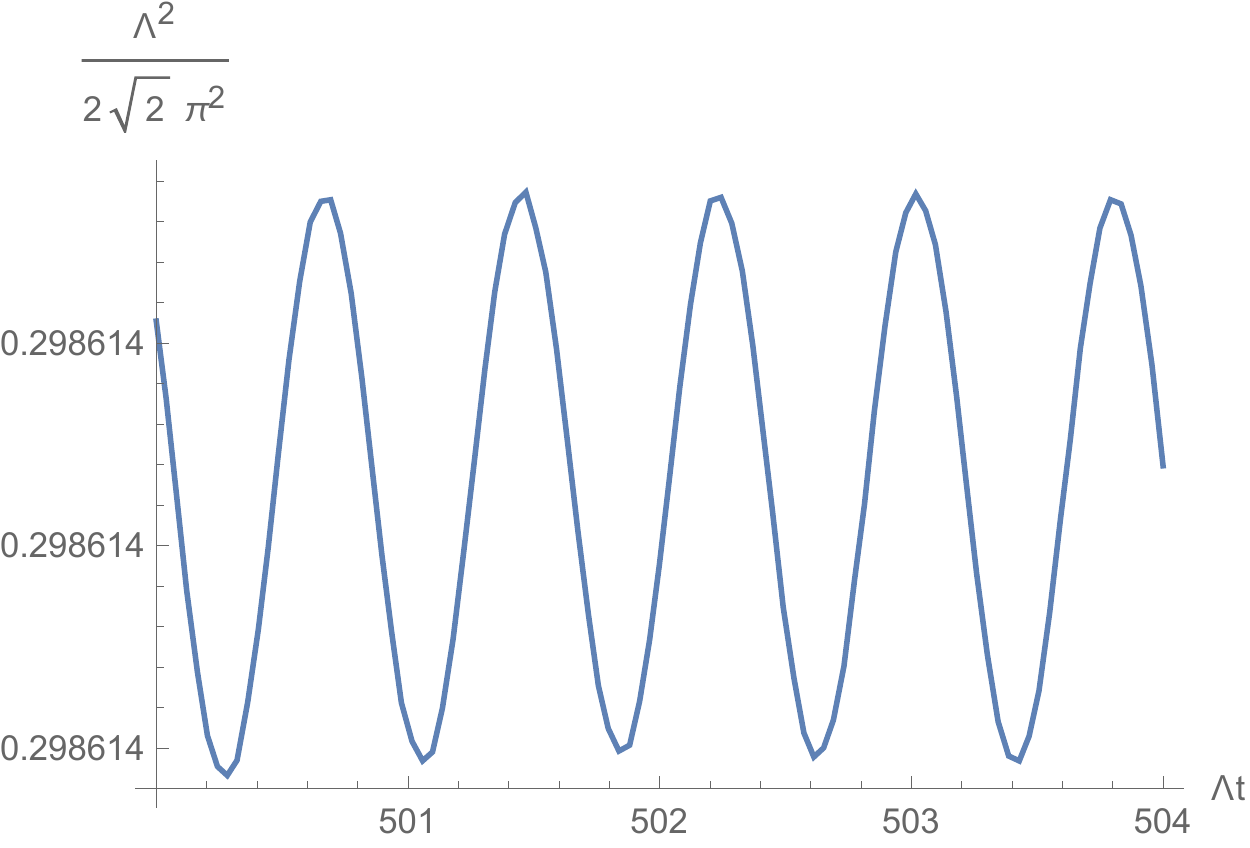}
        \caption{Oscillation of $G_R(0,t)$, when $t$ is large.}
        \label{fig:my_label}
    \end{figure}
    Now, a problem arises naturally: How to interpret this result. In Fig.1 and Fig.2, the modified Green's function starts at the origin and finally reaches the peak. Meanwhile after reaching the peak, the function oscillates with a small amplitude. To make a correspondence to the evolution of $\phi$, we notice that when $t=0$  the scalar field $\phi=0$ and then $\phi$ evolves with time. However, $\phi$ cannot always increase. This is because once it starts evolving not at the true vacuum - the global minimal point of the potential, there is always a tendency get back to the stable true vacuum. Taking this thought into account, $G_R(0,0)=0$ indicates that the scalar field $\phi$ starts at the origin and reaches the true vacuum periodically. Because the energy for driving the inflation of the Universe is gradually dissipated, the amplitude of $\phi$ becomes smaller and smaller. Finally,  $\phi$ oscillates around the global minimal point with a small amplitude. This idea can be true only when we admit such a postulation:
    \begin{align}
        \text{The maximum of } G_R\approx\frac{(2-\sqrt{2})\Lambda^2}{8\pi^2}=v=\frac{6\mu^2}{\lambda}.
    \end{align}
    Therefore, we obtained a correlation of the ultraviolet cutoff $\Lambda$ and the global minimal of the potential.This kind of relationship does not appear in QFT. This is because in QFT, the ultraviolet cutoff is set by an exterior constant which is there just for regularization. However, for this case, we hope to establish an effective field theory which takes into considerations of the quantum vacuum fluctuations. In this case, the key question is how the energy density of this effective field drives the Universe to expand. As a result, there is an exterior constraint on the cosmological background and we cannot treat it as the usual QFT in fixed background.

\section{\label{sec:level1}Toy model: How the inhomogeneous vacuum influences the evolution of the Universe}
\renewcommand{\thefootnote}{\alph{footnote}}

After deriving the modified Green's function, now we are able to explore how the inhomogeneous vacuum influences the evolution of the Universe. Recall that one of the Einstein's equation (15) in section II, adopting the approximation in (41) and substituting it into (15), and then we have\footnotemark[1]
\begin{align}
    -\frac{3\ddot{a}}{a}=8\pi G(\langle \dot{\phi}\dot{\phi}\rangle+\frac{\mu^2}{2}G_R+\frac{\lambda}{4!}{G_R}^2+\frac{3\mu^2}{2\lambda})
\end{align}
However, here, a new issue emerges, that is how to calculate $\langle\dot{\phi}\dot{\phi}\rangle$. As a matter of fact, we have such a relation between $\langle\dot{\phi}\dot{\phi}\rangle$ and the modified Green's function
\begin{align}
    \langle\dot{\phi}\dot{\phi}\rangle(x)=\lim_{x\rightarrow x'}\langle\dot{\phi}(x)\dot{\phi}(x')\rangle=\lim_{x\rightarrow x'}\partial_{t}\partial'_{t}\langle\phi(x)\phi(x')\rangle.
\end{align}

In analogy to the method that we use to derive the modified Green's function for $\langle\phi\phi\rangle$ in the previous section, functions $A(x,t)$ and $B(x,t)$ can also be introduced, and due to (45), the new $A(x,t)$ and $B(x,t)$ are given as
\begin{align}
    A'(\mathbf{x},t)\!=\!&\int \frac{d^3kd^3k'}{(2\pi)^6}\frac{\sqrt{\omega\omega'}}{2}(a_k a_{k'}\!+\!a^\dagger_k a^\dagger_{k'})\cos(k\!+\!k')x,\nonumber \\
    B'(\mathbf{x},t)\!=\!&\int \frac{d^3kd^3k'}{(2\pi)^6}\frac{\sqrt{\omega\omega'}}{2}(a^\dagger_k  a^\dagger_{k'}\!-\!a_k a_{k'})\sin(k\!+\!k')x.
\end{align}
After the direct calculations, results in (46) are quite similar to that of the $G_R$. Therefore, the regulation is also necessary for $\langle\dot{\phi}\dot{\phi}\rangle$. However, unlike the method we used for the modified Green's function, the final version of $\langle\dot{\phi}\dot{\phi}\rangle$ after regulation becomes
\begin{align}
    \langle\dot{\phi}\dot{\phi}\rangle=&\sqrt{\langle A'^2(\mathbf{x},t)\rangle}+\sqrt{\langle B'^2(\mathbf{x},t)\rangle}\nonumber\\&-\lim_{t\to\infty}\left(\sqrt{\langle A'^2(\mathbf{x},t)\rangle}+\sqrt{\langle B'^2(\mathbf{x},t)\rangle}\right).
\end{align}
The reason that this type of regulation is adopted, which is different as what we used for $G_R$, is not hard to explain. It is clear that the scalar field $\phi$ finally decays to the true vacuum which is the minimal of the potential energy. Equivalently, recalling the interpretation for $G_R$ in the previous section, we can discuss the correspondence to the evolution of $\phi$. We notice that when $t$ grows larger and larger, $\phi$ will be stable at the minimal point.  Meanwhile, all the derivatives of $\phi$ should be approximately zero at this stage in order to preserve the stability.
\begin{figure}[t]
    \includegraphics[width=0.4\textwidth]{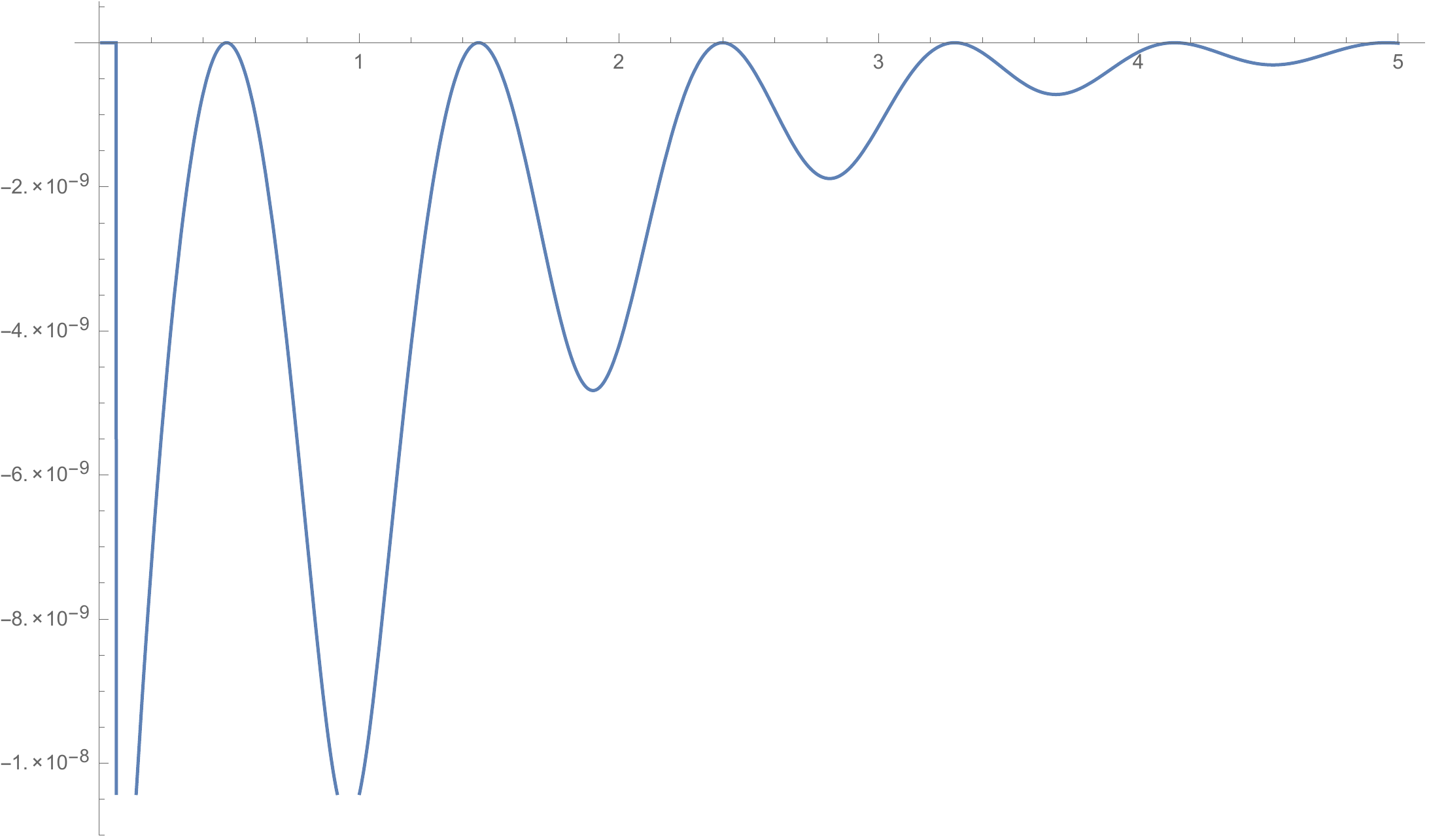}
    \caption{The correlation function $\langle\dot{\phi}\dot{\phi}\rangle$ after regulated in (47)}
    \label{fig:my_label}
\end{figure}

Substituting (47) into (44), we are able to solve the equation for the evolution of the scale factor $a$. After numerical calculation, two figures are plotted for the scale factor $a(0,t)$ comparing to the standard scenario.
\begin{figure}[t]
    \includegraphics[width=0.4\textwidth]{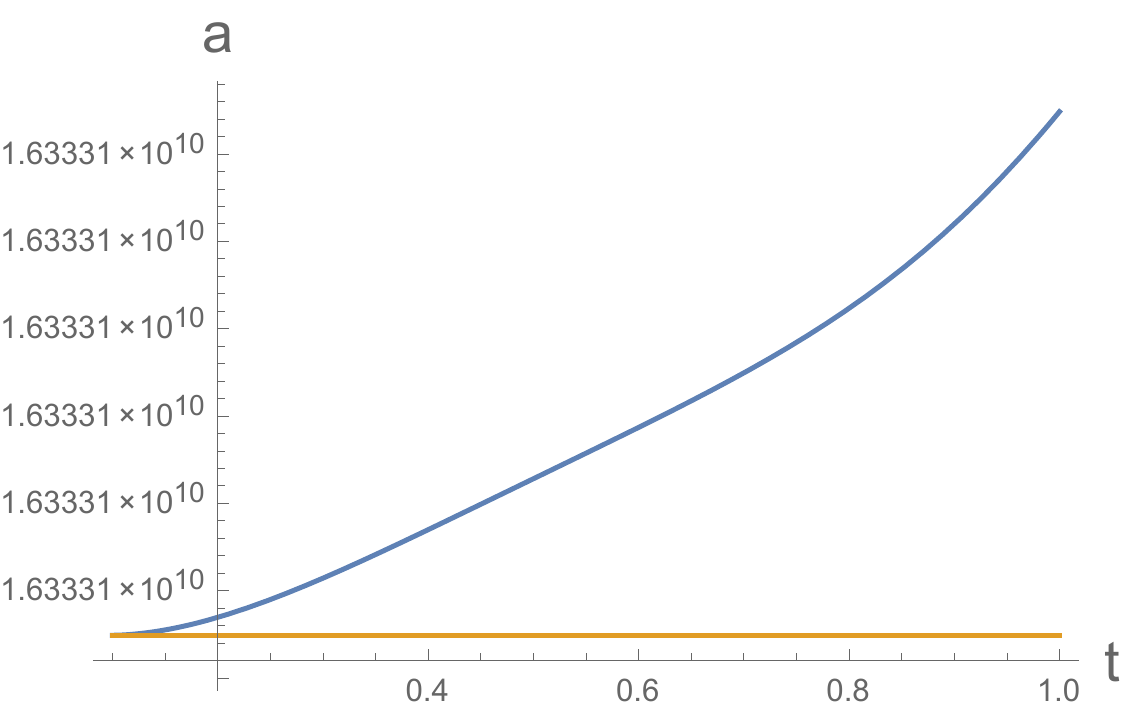}
    \caption{Scale factors at the beginning. The blue curve is for our toy model, and the orange one is for standard scenario.}
    \label{fig:my_label}
\end{figure}
\begin{figure}[t]
    \includegraphics[width=0.4\textwidth]{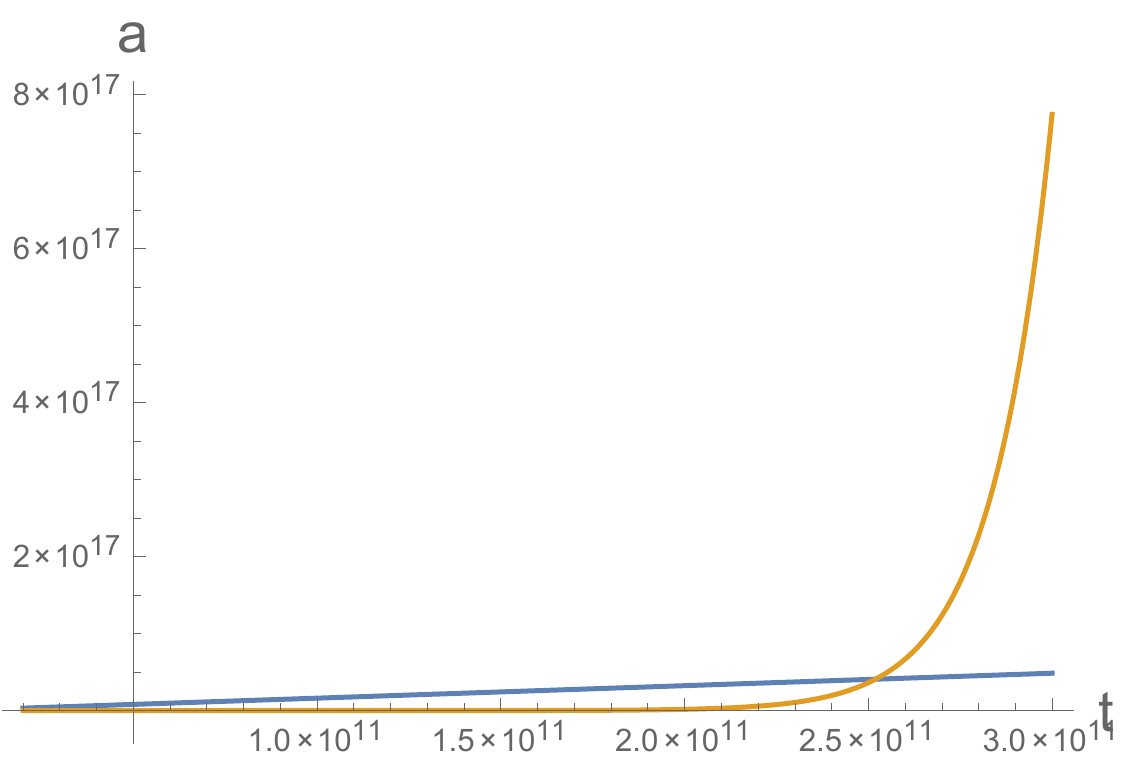}
    \caption{Scale factors when time is large. The blue curve is for our toy model, and the orange one is for standard scenario.}
    \label{fig:my_label}
\end{figure}

In Fig.5 and Fig.6, it is not hard to see that at the beginning, the scale factor in our toy model grows much faster that the scale factor in the standard scenario. However, when the time becomes larger, the scale factor in our toy model grows slower and finally $a\propto t$. This is because when $t$ is large, the scalar field $\phi$ falls down from the top of the potential and oscillates around the global minimal point. In this case, the right hand side of (44) is approximately zero, so $\dot{a}$ is also approximately zero.\footnotemark[1]
\footnotetext[1]{In fact, here, $\phi^4$ should be replaced by 4-point correlation function, which includes at least one four-point interaction. However, once $O(\lambda)$ and higher order diagrams are omitted, ${G_R}^2$ is still good enough to be used for the approximation.}

\section{\label{sec:level1}More elaborated model: When temperature involves}
Once the temperature is introduced in our model, there are significant changes. Based on the finite temperature quantum field theory, correlation functions should satisfy Kubo-Martin-Schwinger relations\cite{ashok}
\begin{equation}
   \langle \chi(t)\chi(t')\rangle_\beta=\langle \chi(t')\chi(t+i\beta)\rangle_\beta.
\end{equation}
Here, time $t$ is also extended to the complex plane and $\beta$ in (48) is the reciprocal for the temperature, $\beta=T^{-1}$. Not only the background becomes more complicated, but the Lagrangian also changes. If the field is not at zero Kelvin, the effective mass becomes
\begin{align}
    \mu^2\to \mu^2-\frac{\lambda}{4!}T^2.
\end{align}
Thus, the Lagrangian becomes
\begin{align}
    \mathscr{L}= \frac{1}{2}g^{\mu\nu}\partial_\mu \phi \partial_\nu \phi+\frac{1}{2}\left(\mu^2-\frac{\lambda}{4!}T^2\right)\phi^2-\frac{\lambda}{4!}\phi^4-\frac{3\mu^4}{2\lambda},
\end{align}
and in this case, the minimal position of the potential is no longer at $\nu=6\mu^2/\lambda$ but at
\begin{align}
    \nu(T)=\frac{6\mu^2(T)}{\lambda}=\frac{6}{\lambda}\left(\mu^2-\frac{\lambda}{4!}T^2\right)=\nu(0)-\frac{T^2}{4}.
\end{align}
Recalling the postulation in (43), the ultraviolet cutoff now in finite temperature field theory is related to the temperature $\Lambda=\Lambda(T)$. But, what would happen when  the temperature is higher than the critical point $T>T_c=\sqrt{24\mu^2/\lambda}$? In this case, $\nu(T)=0$, and the spontaneous symmetry breaking no longer exists. Meanwhile, due to the postulation in (43), the ultraviolet cutoff should be zero. However, the behavior of $G_R$ when time becomes long gives us hint that $\phi$ should oscillate around the origin at the minimal of the potential. For more details, the modified Green's function should be discussed, which involves temperature. Similar to the zero temperature approach, we start with $A(x,t)$ and $B(x,t)$. Replacing $t\to t+i\beta$, we reach \footnotemark[2]
\begin{align}
    A(\mathbf{x}\!,\!t)\!=\!&\!\int\! \frac{d^3kd^3k'}{2(2\pi)^6\sqrt{\omega\omega'}}\!(a_k a_{k'}e^{\beta}\!+\!a^\dagger_k a^\dagger_{k'}e^{-\beta})\cos(k\!+\!k')x,\nonumber \\
    B(\mathbf{x}\!,\!t)\!=\!&\!\int\! \frac{id^3kd^3k'}{2(2\pi)^6\sqrt{\omega\omega'}}\!(a^\dagger_k  a^\dagger_{k'}e^{-\beta}\!-\!a_k a_{k'}e^{\beta})\sin(k\!+\!k')x.
\end{align}
Noticing here, the square of the effective mass in $\omega$ is $m_{eff}^2=\lambda T^2/4!-\mu^2$. Following the process in previous section, $\sqrt{|A(x,t)|^2}$ and $\sqrt{|B(x,t)|^2}$ need to be calculated. Nevertheless, the term which depends on $\beta$ is canceled out. Therefore, finally, the modified Green's function here is no different from the previous one. To find the explicit form when $T$ is very high, considering the approximation in (39) and $\Lambda<<\lambda T^2$, we reach
\begin{align}
     G_R(0,t)\approx \frac{\sqrt{2}}{12\pi^2}\frac{\Lambda^3}{m_e}(|\cos2m_{e}t|+|\sin2m_{e}t|-1).
\end{align}
where $m_{e}^2=\lambda T^2/4!-\mu^2\approx \lambda T^2/4!$. It is easy to find out that $G_R(0,t)$  oscillates at the origin and the amplitude is small.

Before solving the Einstein's field equation for the scale factor, we notice that the term $\langle\dot{\phi}\dot{\phi}\rangle$ in (15) is still unknown. Similar to the previous described approach with zero temperature, introducing $A'(x,t)$ and $B'(x,t)$ in (46) for the finite temperature field, we reach
\begin{align}
    A'(\mathbf{x}\!,\!t)\!=\!&\!\int\!\frac{d^3kd^3k'}{(2\pi)^6}\!\frac{\sqrt{\omega\omega'}}{2}\!(a_k a_{k'}e^{\beta}\!+\!a^\dagger_k a^\dagger_{k'}e^{\!-\!\beta})cos(k\!+\!k')x,\nonumber \\
    B'(\mathbf{x}\!,\!t)\!=\!&\!\int\! \frac{d^3kd^3k'}{(2\pi)^6}\!\frac{\sqrt{\omega\omega'}}{2}\!(a^\dagger_k  a^\dagger_{k'}e^{\!-\!\beta}\!-\!a_k a_{k'}e^{\beta})sin(k\!+\!k')x.
\end{align}
Using the approximation $\omega\approx m_{eff}$, in this case, we reach
\begin{align}
    \langle\dot{\phi}\dot{\phi}\rangle=m_{eff}^2G_R.
\end{align}
Then plugging (53) and (55) into (15), we obtain the equation for the scale factor
\begin{align}
    \frac{\ddot{a}}{a}=\frac{8\pi G}{3}(-\frac{1}{2}m_{eff}^2(T)G_R+\frac{\lambda}{4!}G_R^2+\frac{3\mu^4}{2\lambda}).
\end{align}
To solve this equation, the relation between the temperature and the scale factor is needed. Due to the total entropy conservation, the relation is given as $T\propto a^{-1}$. In (56), because the amplitude of $G_R$ is negligibly small, the last term $V(0)=(3\mu^4)/(2\lambda)$ dominates the equation. Therefore, it is not hard to predict that the scale factor $a$ increases exponentially before the temperature is below the critical temperature.
\footnotetext[2]{Strictly speaking, the imaginary time method should be applied to discuss the modification of $A$ and $B$, but it makes no difference to our result. For details of the imaginary time, see Ref.\cite{ashok}.}

\section{\label{sec:level1}A new inflationary scenario}
In this section, we mainly discuss a new inflationary scenario which is built based on our model. At the beginning, in this new inflationary scenario, we investigate the idea that the Universe is created from nothing\cite{vilenkin,vilenkin2}. Hence, in our model, the curvature is greater than zero, or equivalently, we take $k=1$ in the FLRW metric. From the tunneling theory, before tunneling through the barrier, the 'Universe' was in Euclidean space. After penetrating the potential barrier, the Universe was 'created'
and began expanding. At this stage, the vacuum energy drove the Universe expanding exponentially. Following the previous section, the symmetry was restored because the field was at a very high temperature which is much higher than the critical temperature where phase transition of the field occurs. Therefore,  $\phi$ oscillates at the origin with a tiny amplitude which can almost be negligible. By omitting the terms containing $G_R$ in (56), during this period, the equation for the scale factor becomes
\begin{align}
    \frac{\ddot{a}}{a}\approx \frac{8\pi G}{3}\frac{3\mu^4}{2\lambda}=\frac{4\pi G\mu^4}{\lambda}=H^2.
\end{align}
Take the initial condition when $t\approx0$, we reach  $a(x,t)=a(t)= H^{-1}cosh(Ht)$. This also gives the solution to the Friedmann equation for homogeneous and isotropy space-time with positive curvature\cite{vilenkin2}. In this case, we have the same solution to (57)
\begin{align}
    a\approx H^{-1}\cosh(Ht).
\end{align}
In fact, following the Friedmann equations, besides the 00 component of Einstein's field equation as (15), the other ones are derived from the first one and the trace of Einstein's field equations. Similarly, we can derive the second equation for our model.
\begin{widetext}
\begin{align}
    \mathbf{LHS.}&=\frac{1}{6}\left(\frac{R_{00}}{g_{00}}-\frac{R_{11}}{g_{11}}-\frac{R_{22}}{g_{22}}-\frac{R_{33}}{g_{33}}\right)=\frac{4 \left(\left(3 r^2-2\right) a'+r \left(r^2-1\right) a''\right) a-2r \left(r^2-1\right) a'^2}{6r a^4}+\frac{\dot{a}^2+1}{a^2},\nonumber\\
    \mathbf{RHS.}&=\frac{1}{6}\left(\frac{T_{00}}{g_{00}}-\frac{T_{11}}{g_{11}}-\frac{T_{22}}{g_{22}}-\frac{T_{33}}{g_{33}}\right)=\frac{8\pi G}{6}\left(\langle\dot{\phi}\dot{\phi}\rangle+\frac{1-r^2}{a^2}\langle\phi'\phi'\rangle+2V(\phi)\right)\approx\frac{8\pi G}{3}V(0)=H^2.
\end{align}
\end{widetext}
On the left hand side of the second Friedmann equation, it separates into two parts. One part comes from the spatial dependence of the scale factor, and the other part is a common term. Because the right hand side is approximately a constant, we can also make the approximation so that $a(r,t)\approx a(t)$. Then the first part on the left hand side disappears. Hence, a much more simplified equation is derived
\begin{align}
    \dot{a}^2+1\approx H^2a^2.
\end{align}
It is obviously that the solution to this equation is the same as to that of (57).

At the first stage, the Universe expanded exponentially, so the temperature of the entire Universe decreased rapidly, due to the relation of temperature and entropy\cite{guth}, $T\propto s^{-1}$. Once the temperature becomes lower than the critical temperature, the Universe switched to the next stage. Recalling the section IV, and considering $\langle\dot{\phi}\dot{\phi}\rangle$ goes to zero and $G_R$ reaches the stable point extremely fast, these terms make very limited contributions. Therefore, they can be neglected in the large scale space-time. However, if we want to solve the second Friedmann equation, (59), for the scale factor during this period, there is still an issue how the spatial dependence of the scale factor influences the result. Ideally, if the spatial factor has no significant effects, we can also drop those terms involving $a'$ and $a''$, and obtain a simple equation as (60). To check this assumption, both two variables should be considered for $G_R$. To illustrate how the spatial part influences the modified Green's function, we plot a 3D figure for $G_R(r,t)$.
\begin{figure}[t]
    \includegraphics[width=0.4\textwidth]{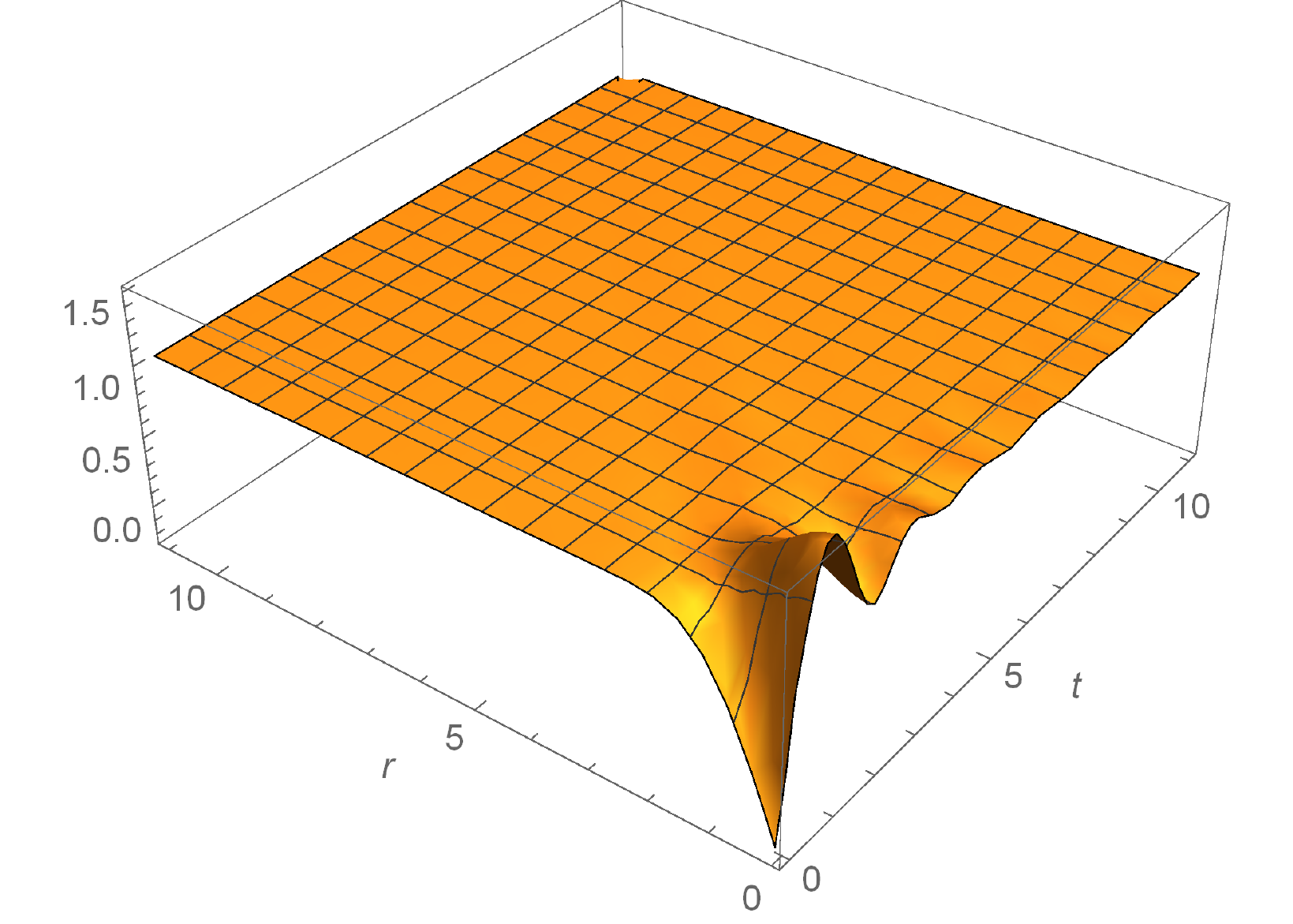}
    \caption{Numerical result of $G_R(r,t)$ in 3D.}
    \label{fig:my_label}
\end{figure}
In Fig.7, except the very little area where $t$ is small, the surface in the figure is almost flat. This means that the spatial fluctuations of the variables are very small that can be omitted when $t$ is not too small. Thus, it is fine to get rid of the part which contains $a'$ or $a''$ in (59), and obtain a new equation
\begin{align}
    \dot{a}^2+1\approx \frac{8\pi G}{3}a^2V(\phi,T)\Big|_{min}\!=\!\frac{8\pi G}{3}\left(\frac{c^2T^2}{8}-\frac{\lambda T^4}{384}\right)a^2.
\end{align}
Because the expansion of the Universe is adiabatic, so the total entropy of the Universe is conserved and set $aT=S$, then $T$ can be replaced in (61)
\begin{align}
    \dot{a}^2+1 &\approx \frac{8\pi G}{3}\left(\frac{\mu^2S^2}{8a^2}-\frac{\lambda S^4}{384a^4}\right)a^2.
\end{align}
In this switching period, (62) could be rewritten as
\begin{align}
    \dot{a}&\approx \left[\left(\frac{\pi G\mu^2S^2}{3a^2}-1\right)-\frac{\pi G\lambda S^4}{144a^2}\right]^{1/2}\nonumber\\&=\sqrt{P-Q/a^2}.
\end{align}
So, it is easy to find that when scale factor grew larger, the second term Q/a2 became smaller and $a'\approx Constant$, which means the Universe expanded approximately in a fixed speed at this stage. In this case, the expansion of the Universe was linearly in time. Meanwhile, during this period, the acceleration of the expansion kept on decreasing (In fact, the acceleration cannot be zero but at a very small value, we will explain this effect next.)

After the switching period, the universe cooled down. due to the coupling with other fields, the bosons of the scalar field would decay into other particles. Thus, the $P=P(t)$ and $Q=Q(t)$ in (63) dropped down, so the speed of linear expansion in the last period also decreased rapidly. Moreover, at this stage, the vacuum energy of the scalar field contributed to the Universe's expansion decreases, so the speed of expansion continued decreasing. But it does not mean the Universe stopped expanding. In fact, the expansion of the Universe was still accelerating because of one element that we neglect in the previous stage, which became more important. In section III, the behavior of $G_R$ shows that the field $\phi$ falls into the minimal  of the potential very fast. But once it reaches that point, it will oscillate at that point with a tiny amplitude. The oscillation makes a tiny shift, which is quite small comparing to the right hand side of (61). To calculate the tiny shift caused by the oscillations of $\phi$, the two approximations in section III which are applied to calculate $G_R$ should be both considered, when $t$ is large, we reach
\begin{widetext}
\begin{align}
    G_R(\mathbf{x},t)&=\sqrt{A_1^2(\mathbf{x},t)+A_2^2(\mathbf{x},t)}+\sqrt{B_1^2(\mathbf{x},t)+B_2^2(\mathbf{x},t)}-\sqrt{A_1^2(0,0)+A_2^2(0,0)}-\sqrt{B_1^2(0,0)+B_2^2(0,0)}\nonumber\\&\approx \left(\sqrt{A_1^2(\mathbf{x},t)}+\sqrt{B_1^2(\mathbf{x},t)}-\sqrt{A_1^2(0,0)}-\sqrt{B_1^2(0,0)}\right)+\frac{1}{2}\left(\frac{A_2^2(\mathbf{x},t)}{\sqrt{A_1^2(\mathbf{x},t)}}+\frac{B_2^2(\mathbf{x},t)}{\sqrt{B_1^2(\mathbf{x},t)}}\right)\nonumber\\&\approx \nu(T)+\frac{1}{36\pi^2}\frac{{\Lambda_0}^6}{\Lambda^2(T)\mu^2(T)}.
\end{align}
\end{widetext}
where $\Lambda(T)$ is the ultarviolet cutoff which is related to temperature.  $\mu^2(T)=\mu^2-\lambda T^2/4!$ is the square of the effective mass in this temperature.  $\nu(T)=6\mu^2(T)/\lambda$ indicates the minimal point of the potential. Thus, we can define the shift of the scalar field $\phi$
\begin{align}
    (\Delta\phi)^2=\lim_{t\to\infty}G_R(\mathbf{x},t)-\nu(T)=\frac{1}{36\pi^2}\frac{{\Lambda_0}^6}{\Lambda^2(T)\mu^2(T)}.
\end{align}
Then, the tiny shift of the potential can also be obtained
\begin{align}
    \Delta V=V(\phi_0+\Delta\phi)-V(\phi_0)\approx\frac{V''(\phi_0)}{2}(\Delta\phi)^2.
\end{align}
where $\phi_0=\sqrt{\nu(T)}$, which indicates the minimal point, and $V''=\partial^2 V/\partial\phi^2$. Substituting (65) and the Lagrangian into (66), here we have
\begin{align}
    \Delta V=\frac{{\Lambda_0}^6}{36\pi^2\Lambda^2(T)}.
\end{align}
Currently, in the case that $T\approx 2.7k$, we have $\Lambda^2(T)\approx\Lambda^2(0)\approx(48\pi^2\mu^2)/((2-\sqrt{2})\lambda)>>\Lambda_0$, so the energy shift $\Delta V$ is a very small constant.

Besides the energy shift began playing a role in the Universe expansion, after the particles of the scalar field decaying into other particles, the density of particle $\phi$ kept decreasing. According to the radiation thermodynamics, the density of particle is proportion to $T^4$.\cite{weinberg} Therefore, in this period, the particle density is given as
\begin{align}
    \rho=\rho_0e^{-\Gamma t}\propto T^4e^{-\Gamma t}.
\end{align}
where $\Gamma$ is the decay rate which is dependent on the coupling constants of $\phi$ and other fields. Therefore,  equivalently, we can replace the original temperature by $T\to Texp(-\Gamma t/4)$. Now, we are able to establish the equation for the last stage of the inflation evolution
\begin{align}
    \dot{a}^2\!+\!1\approx \frac{8\pi G}{3}\!\left(\!\frac{\mu^2S^2}{8a^2}e^{-\Gamma t/2}\!-\!\frac{\lambda S^4}{384a^4}e^{-\Gamma t}\!+\!\frac{\rho}{a^3}\!+\!\Delta V\!\right)\!a^2.
\end{align}
where $\rho$ indicates the density of non-relativistic matter. Now, we have established the equation which can explain the scenario of the universe nowadays. In this equation, the first term which is proportional to $a^{-2}$ decreases rapidly. The next term is negligible when $t$ is large. The third term comes from the non-relativistic matter which is created by the energy that the scalar field dissipates. The last term, which is approximately a constant, plays the role of vacuum energy or, in another word, the cosmological constant.

According to the observations, now, the universe is still expanding, and the expansion is accelerating. Using the observational data, one can check which factor dominates the current Universe evolution. To describe the behavior of the current Universe, a general evolution equation is introduced
\begin{align}
    \dot{a}^2=H_0^2\left[\Omega_{\Lambda}+\Omega_K\left(\frac{a_0}{a}\right)^2+\Omega_M\left(\frac{a_0}{a}\right)^3+\Omega_R\left(\frac{a_0}{a}\right)^4\right].
\end{align}
where $H_0$ is the current Hubble's parameter, $a_0$ is the current scale factor. Meanwhile, the energy densities of non-relative matter, radiation, and vacuum are respectively
\begin{align}
    \rho_{M0}=\frac{3H_0^2\Omega_M}{8\pi G},\rho_{R0}=\frac{3H_0^2\Omega_R}{8\pi G},\rho_{\Lambda0}=\frac{3H_0^2\Omega_{\Lambda}}{8\pi G}.
\end{align}
By fitting with the observed data of the supernova\cite{Riess,Riess2}, the ratio of each component which sustains the expansion of the Universe can be obtained. One of  results obtained is that $\Omega_K=\Omega_R=0$,  $\Omega_{\Lambda}+\Omega_M=1$, and $\Omega_{\Lambda}\approx 0.7$. In (69), because at this stage, $t$ is large, so the first term is small. To match the current statue of the Universe, an approximate formula can be laid down.
\begin{align}
    \frac{c^2S^{2/3}}{8}e^{-\Gamma t/2}\approx\frac{3}{8\pi G}.
\end{align}
Thus, the curvature constant is effectively equal to zero. Then, by neglecting the second term on the right hand side of (69), we reach the equation which describes the current Universe as
\begin{align}
    \dot{a}^2\approx \frac{8\pi G}{3}\left(\frac{\rho}{a^3}+\Delta V\right)a^2.
\end{align}
In this case, the expansion of the universe is dominated by vacuum energy which is from the energy shift caused by the oscillations of the scalar field $\phi$ around the potential minimal and non-relativistic matter, while other effects are ruled out. Meanwhile, plugging the current observed Hubble constant\cite{Hubble} in (71), we obtain the energy density of vacuum energy (cosmological constant):
\begin{align}
    \rho_{\Lambda 0}\approx2.57\times 10^{-47}(GeV)^4.
\end{align}
This result is much smaller than the energy scales of various theories of QFT.  Since $\Delta V$ in (67) is also much smaller than the energy that QFT can predict, this result can be used to explain why the current energy density of the vacuum energy is so small \cite{Unruh}. In our model, the density of effective vacuum energy caused by $\Delta V$ is
\begin{align}
    \rho_{\Lambda}=\Delta V \propto \Lambda^{-2}.
\end{align}
In this case, the Hubble parameter $H\propto\Lambda^{-1}\to 0$. Comparing to another resolution to the cosmological constant problem \cite{Unruh}, the effective Hubble parameter approaches to zero, $H\propto\Lambda e^{-\beta\sqrt{G}\Lambda}\to 0$. Although both two models draw a conclusion that the Hubble parameter $H\to 0$ when the ultraviolet cutoff is taken as infinity, $\Lambda\to \infty$, the results in the two models are still different. This is because in our model, the quantum vacuum fluctuations have direct impacts on the scalar field $\phi$. Thus we obtain the effective density of vacuum energy $\Delta V$.  In reference \cite{Unruh}, the vacuum fluctuations have direct impacts on the space-time structure, but not on the field itself.
\begin{figure}
        \centering
        \includegraphics[width=0.45\textwidth]{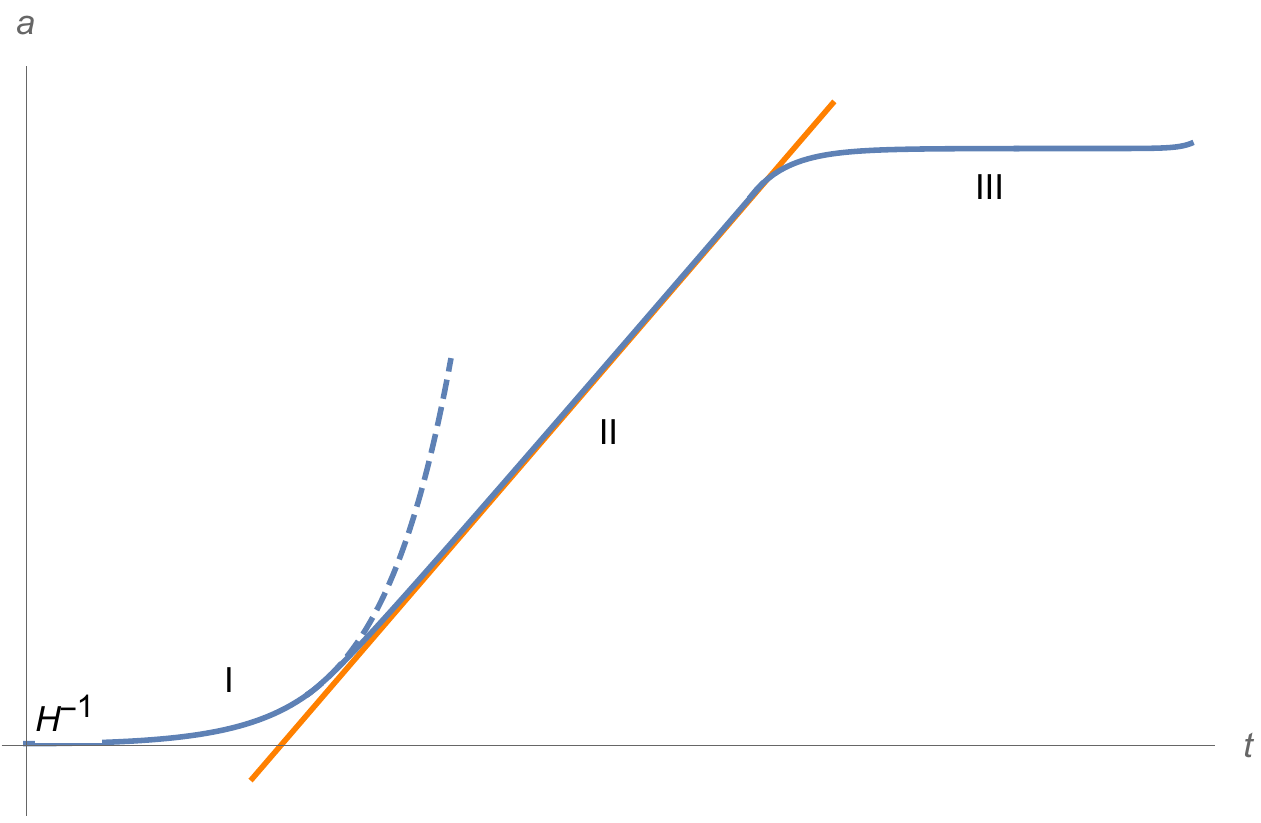}
        \caption{The scale factor of the new scenario. Notice there are three stages of the evolution of the scale factor.}
        \label{fig:my_label}
\end{figure}
To summarize, there are three main stages of
this new inflation scenario (shown in Fig.8): At the first stage (the curve I in Fig.8), the scalar
field had no symmetry breaking at the very high
temperature and the Universe expanded exponentially. Meanwhile, the temperature kept decreasing rapidly and once it dropped blow the
critical point, the scenario switched to the next stage. At this transient period (the curve II in Fig.8), the expansion was no longer exponential but the acceleration of the expansion decreased, and finally the expansion was approximately linearly in time. Next, along the decrease of the temperature, the $P$ and $Q$ in (63) can no longer be as constants anymore but went down
quickly. Therefore,  in this stage (the curve III in Fig.8), the speed of the expansion became slower and slower, and finally $P(t)\approx0$. However, one effect buried in previous stages became important now. After $P(t)$ being approximately zero, the effective potential in (67) leads to the acceleration of the expansion of the current
Universe.

Comparing to the standard inflationary scenario, the new scenario proposed here has no issue of how the Universe quit the inflationary stage from the exponential expansion. The acceleration of expansion decreased smoothly from the first stage to the second stage, and finally the expansion was almost linearly in time. This
smooth transition is guaranteed by the behavior
of the modified Green’s function which shows that the scalar field at each point of the whole Universe can reach the local minimum on the effective potential landscape. Therefore, the new inflationary scenario does not have to quit through bubble collisions.

\section{\label{sec:level1}Tunneling amplitude: How  the inhomogeneous vacuum influences the creation of the Universe}
After establishing the model describing the evolution of the universe, let us go back to the moment where the Universe has not been created. Due to the tunneling  theory\cite{vilenkin,vilenkin2}, the universe can be created from 'nothing' by tunneling through the potential barrier. It is therefore useful to estimate the tunneling amplitude in our model, and the influence of the inhomogeneous vacuum on the tunneling amplitude. To compute the amplitude, we followed the following step. First find the minimal coupling action. Then derive the Wheeler-Dewit (W-D) equation. Finally solve the W-D equation and then find the outgoing wavefunction and calculate the tunneling amplitude.

The minimal coupling of Einstein-Hilbert action and the scalar field action is given as
\begin{equation}
   S=\int \left[-\frac{R}{16\pi G}+\frac{1}{2}g^{\mu\nu}\partial_\mu \phi \partial_\nu \phi-V(\phi) \right]\sqrt{-g}d^4 x,
\end{equation}
where the potential $V(\phi)$, based on our model, is given as
\begin{equation}
   V(\phi)=\frac{1}{2}\mu^2\phi^2+\frac{1}{4!}\lambda\phi^4+\frac{3\mu^4}{2\lambda}.
\end{equation}
By setting $r=sin\chi$, the metric becomes $ds^2=dt^2-a^2(t,\chi)[d\chi^2+sin^2\chi(d\theta^2+sin^2\theta d\phi^2)]=dt^2-a^2(t,\chi)d\Omega^2_3$. Then after direct calculation, the Ricci scalar in our model is given as
\begin{equation}
    R\!=\!-\frac{2}{a^4} \left[a'^2\!-\!2a\left(2
    \cot(\chi)a'\!+\!a''\right)\!+\!3(1\!+\!\dot{a}^2)\!+\!3a^3\ddot{a}\right],
\end{equation}
where $\dot{a}=\partial a/\partial t$ and $a'=\partial a/\partial \chi$. Substituting the potential and Ricci scalar into action and writing out the action explicitly, we have
\begin{equation}
   S=\int dt\ \mathrm{L}(a,\dot{a},a';\phi,\dot{\phi},\phi')
\end{equation}
where the Lagrangian reads
\begin{widetext}
\begin{align}
   \mathrm{L}=\int &\left[\frac{1}{8\pi Ga^4}(a'^2-2a\left(2\cot(\chi)a'+a''\right)+3(1+\dot{a}^2)a^2+3a^3\ddot{a})+\dot{\phi}^2-\frac{\phi'^2}{a^2}-V(\phi) \right]a^3sin^2\chi\sin\theta d\chi d\theta d\varphi.
\end{align}
\end{widetext}

There is an issue involving the difficulty in evaluating the integrations in (80). To evaluate the integral, we make the approximation that $\phi$ and $a$ are approximately constants when integrating over $\chi$. In fact, this approximation is acceptable. Recalling Fig.7, the spatial variation of $\phi$ is small, and  because $a$ is determined by $\phi$,  the spatial variation of $a$ is also small. Then we have the approximate Lagrangian
\begin{align}
   \mathrm{L}\!=\!2\pi^2\!\left[\!\frac{1}{8\pi Ga}(a'^2\!+\!3(1\!-\!\dot{a}^2)a^2)\!+\!a^3\dot{\phi}^2\!-\!a\phi'^2\!-\!a^3V(\phi)\!\right].
\end{align}
Then, following the same idea, we can make further approximations that neglect $a'$. Therefore, now we have a simpler Lagrangian
\begin{align}
   \mathrm{L}\!=\!2\pi^2\left[\frac{3}{8\pi G}(1\!-\!\dot{a}^2)a\!+\!a^3\dot{\phi}^2\!-\!a\phi'^2\!-\!a^3V(\phi)\right].
\end{align}
Now, we can derive the W-D equation. Firstly, the canonical momenta should be obtained
\begin{equation}
  \left\{
  \begin{aligned}
   P_a&=\frac{\partial L}{\partial a}=\frac{3\pi a\dot{a}}{2G},\\
   P_\phi&=\frac{\partial L}{\partial \phi}=2\pi^2 a^3\dot{\phi}.
\end{aligned}
\right.
\end{equation}
By Legendre transform and substituting the canonical momenta into (82), we have the Hamiltonian
\begin{equation}
  \mathrm{H}\!=\!-\frac{G}{3\pi a}{P_a}^2\!+\!\frac{1}{4\pi^2 a^3}{P_\phi}^2\!-\!\frac{3\pi}{4G}a\left[1\!-\!\frac{8\pi G}{3}\!\left(\!\phi'^2\!+\!a^2V(\phi)\!\right)\!\right].
\end{equation}
To obtain Wheeler-DeWitt equation, introduce canonical quantization by replacing $P_a\rightarrow-i\partial/\partial a$, $P_\phi\rightarrow-i\partial/\partial\phi$, then we have the W-D equation
\begin{align}
   \left[\frac{\partial^2}{\partial a^2}+\frac{p}{a}\frac{\partial}{\partial a}-\frac{1}{a^2}\frac{\partial^2}{\partial \tilde{\phi}^2}-U(a,\phi)\right]\psi=0,
\end{align}
and here, $\tilde{\phi}^2=4\pi G\phi^2/3$, the parameter $p$ represents the ambiguity:
\begin{align}
       {P_a}^2\sim a^2\dot{a}^2=\frac{1}{a^p}(a\dot{a})a^p(a\dot{a}).
\end{align}
In this case, it is not appropriate to just write ${P_a}^2$ as $\partial^2/\partial a^2$, but
\begin{align}
    {P_a}^2=-\frac{1}{a^p}\frac{\partial}{\partial a}(a^p\frac{\partial}{\partial a})=\frac{\partial^2}{\partial a^2}+\frac{p}{a}\frac{\partial}{\partial a}.
\end{align}
In the W-D equation (85), the potential $U$ is given as
\begin{align}
    U(a,\phi)=\left(\frac{3\pi}{2G}\right)^2a^2\left[1-\frac{8\pi G}{3}\left(\phi'^2+a^2V(\phi)\right)\right].
\end{align}

Based on the tunneling approach, before the universe was created, everything was formulated in Euclidean space. To know the behavior of the Universe before tunneling through the potential barrier, gravitational instanton solution should be considered. At this stage, $\phi$ is assumed to not change with time. Therefore, it is possible to set $\phi=0$. This is because in the inflationary scenario that we discussed in the previous section, the scalar field $\phi$ starts at the extremum, $\phi=0$. We assume that before tunneling through the barrier, in Euclidean space, the Universe preserved the highest symmetry. Therefore, we assume that the Universe was homogeneous and the scale factor was not related to $r$. Now we have Freedman equation in Euclidean space as
\begin{equation}
   -\dot{a}^2+1=H^2a^2.
\end{equation}
where $H=(8\pi G V(0)/3)^{1/2}$, $V(0)=\frac{3c^4}{2\lambda}$
The solution to this simple equation is
\begin{equation}
   a(t)=H^{-1}\cos(Ht).
\end{equation}
This solution shows that before the universe 'penetration' through the potential barrier, in Euclidean space, it expanded and contracted periodically. In this case, when we analyze the wave function of the W-D equation, the region $a<H^{-1}$ is in Euclidean space representation. Meanwhile, the region $a>H^{-1}$ is in Lorentz space representation. At the very early time after the universe has been created, $t<<1$, following the Lorentz space representation, we have
\begin{equation}
   \dot{a}^2+1 \approx H^2a^2.
\end{equation}
The solution is
\begin{equation}
   a(r,t)\approx  H^{-1}\cosh(Ht),
\end{equation}
and this is also the initial conditions that we used for our model.

Following the same approach that we obtained the W-D equation for Lorentz region, in $a>H^{-1}$, the W-D equation for Euclidean region is simpler
\begin{align}
      \left[\frac{\partial^2}{\partial a^2}+\frac{p}{a}\frac{\partial}{\partial a}-\frac{1}{a^2}\frac{\partial^2}{\partial \tilde{\phi}^2}-U_0(a,\phi)\right]\psi=0,
\end{align}
where the potential $U_0$ is
\begin{align}
    U_0(a,\phi)=\left(\frac{3\pi}{2G}\right)^2a^2\left[1-H^2a^2\right].
\end{align}
After the W-D equation is obtained, we can calculate the tunneling amplitude, which is proportional to $\psi(H^{-1})/\psi(0)$. WKB approximation can be used to derive the amplitude
\begin{align}
    \frac{\psi(H^{-1})}{\psi(0)}&=exp\left[-\int_0^{H^{-1}} \sqrt{U_0(a)} da\right]\nonumber\\&=exp\left[-\frac{3}{16G^2V(0)}\right]
\end{align}
Because of $\phi'^2$, the potential $U(a)$ is a little bit different than before. In Fig.8, we can see it clearly that there is a clear sharp step at $a=H^{-1}$, which means the tunneling amplitude should be greater than the conventional result $exp(-3/[16G^2V(0)])$. This reflects the influence caused by the inhomogeneous vacuum. Meanwhile, the sharp step  caused by $\phi'^2$ and $\phi$ is also a spatial function. Therefore, the size of the sharp step at $H^{-1}$ is also related to spatial variable $r$. This means that the tunneling probabilities vary at different spatial locations. Therefore, in our model, each spatial point of the universe is not expected to tunnel through the barrier at the same time due to the different tunneling probability.
\begin{figure}
        \centering
        \includegraphics[width=0.45\textwidth]{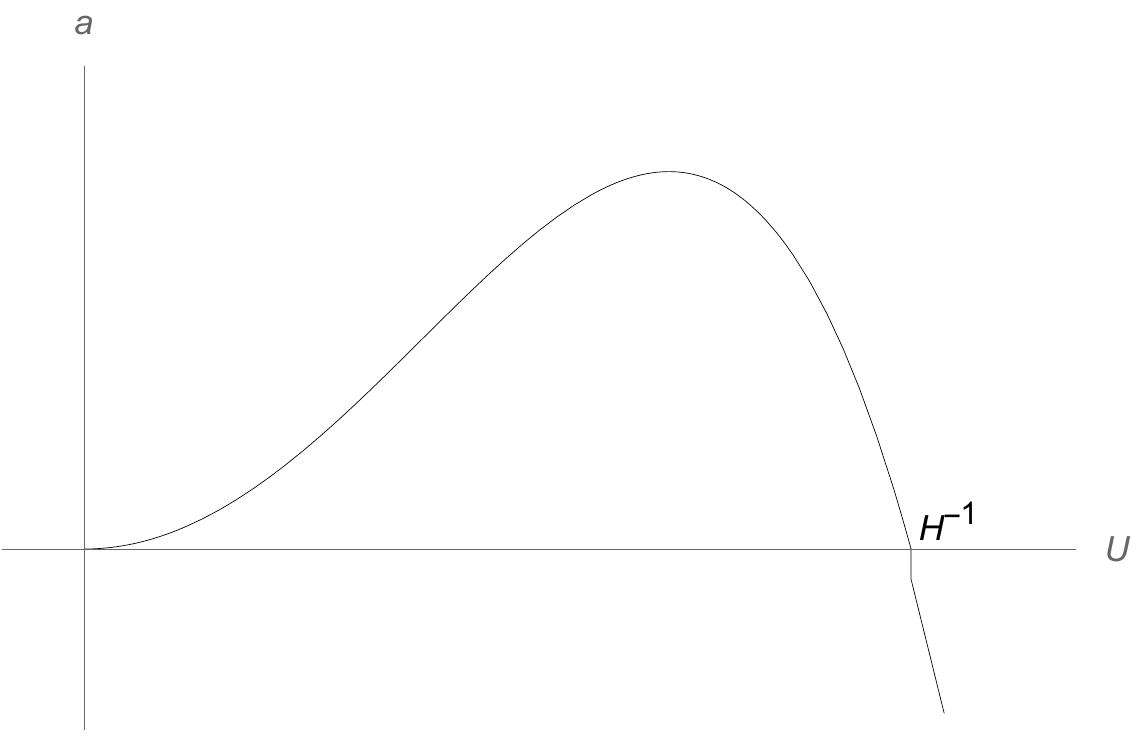}
        \caption{Potential $U(a)$ for W-D equation. Notice the step at $a=H^{-1}$}
        \label{fig:my_label}
\end{figure}
The W-D equation can be solved exactly with the choice of $p=-1$. The choice of the factor-ordering factor $p$ dose not influence the probabilities. Now introducing a new variable
\begin{align}
    z=-(1-a^2H^2).
\end{align}
So the potential $U_0$ becomes
\begin{align}
   U_0=\left(\frac{3\pi}{2G}\right)^2a^2z.
\end{align}
Setting $p=-1$, we have
\begin{align}
   \left[ \frac{\partial^2}{\partial z^2}+\left(\frac{3\pi}{2G}\right)^2\frac{z}{4H^2}\right]\psi=0.
\end{align}
After introducing the new variable, we can rewrite the W-D equation in $a<H^{-1}$. Similarly, we can also introduce another variable for the W-D equation in the region $a>H^{-1}$\cite{vilenkin}
\begin{align}
   z'=-(1-\frac{8\pi G}{3}\phi'^2-a^2H^2).
\end{align}
Because at the very early times, $\phi\approx0$. Therefore, here, $8\pi V(\phi)/3\approx H^2$ and $\phi'^2$ are irrelevant to $a$. The equation can be rewritten once again as
\begin{align}
      \left[ \frac{\partial^2}{\partial z'^2}+\left(\frac{3\pi}{2G}\right)^2\frac{z'}{4H^2}\right]\psi=0.
\end{align}
The solution to these equations are Airy functions
\begin{align}
      \psi=\left\{
      \begin{aligned}
      C_1\mathrm{Ai}(-\sqrt[3]{A}z)+C_2\mathrm{Bi}(-\sqrt[3]{A}z),\ a<H^{-1}\\C'_1\mathrm{Ai}(-\sqrt[3]{A}z')+C'_2\mathrm{Bi}(-\sqrt[3]{A}z'),\ a>H^{-1}.
      \end{aligned}
      \right.
\end{align}
where $A=(3\pi/4HG)^2$. Next, we should find the coefficients in the solution. The continuities of $\psi$ and $\partial \psi/\partial a$ should be considered. Meanwhile, based on the tunneling theory, only an outgoing wave should be considered outside the barrier, which means $i\psi^{-1}\partial \psi/\partial a>0$ for $a>H^{-1}$. For large $z$, we reach the asymptotic formulas for $\mathrm{Ai}$ and $\mathrm{Bi}$
\begin{align}
 \mathrm{Ai}(-z)\sim \frac{\sin \left(\frac23z^{\frac{3}{2}}+\frac{\pi}{4} \right)}{\sqrt\pi\,z^{\frac{1}{4}}}, \nonumber\\
 \mathrm{Bi}(-z){}\sim \frac{\cos \left(\frac23z^{\frac{3}{2}}+\frac{\pi}{4} \right)}{\sqrt\pi\,z^{\frac{1}{4}}}.
\end{align}
Due to the asymptotic formulas and Euler's formula, it is easy to find out that the proper form of $\psi$ for $a>H^{-1}$
\begin{align}
      \psi=C\left(\mathrm{Ai}(-z')+i\mathrm{Bi}(-z')\right).
\end{align}
Then we can obtain the coefficients $C_1$ and $C_2$ related to $C$ by two continuity equations. Now, we are able to calculate the tunneling amplitude
\begin{widetext}
\begin{align}
      \frac{\psi(H^{-1})}{\psi(0)}=\frac{2\cdot3^{3/2}X}{(3^{4/3}\Gamma(\frac23)X-3\Gamma(\frac13)Y) Ai(\sqrt[3]{A})+(3^{5/6}\Gamma(\frac23)X+3^{1/2}\Gamma(\frac13)Y)Bi(\sqrt[3]{A})}.
\end{align}
\end{widetext}
where $X$ and $Y$ inside (104) are 
\begin{align}
    X&=\mathrm{Ai}(-8\pi G\sqrt[3]{A}\phi'^2/3 )+i\mathrm{Bi}(-8\pi G\sqrt[3]{A}\phi'^2/3),\nonumber\\Y&=\mathrm{Ai}'(-8\pi G\sqrt[3]{A}\phi'^2/3 )+i\mathrm{Bi}'(-8\pi G\sqrt[3]{A}\phi'^2/3 ).
\end{align}
Comparing to the conventional solution (95), the result in (104) is much more complicate, and we can see not only $H$ but also $\phi'$ plays an important role in tunneling process, which means the quantum vacuum fluctuation also affects the creation of the Universe.

\section{\label{sec:level1}Conclusion}
In this paper, we analyzed the inhomogeneous vacuum in the Universe, and come up a new method for introducing the inhomogeneity by modifying Green's function. Meanwhile, substituting the modified Green's function in Friedmann equation, we obtained a new inflationary scenario which can explain why the Universe is still expanding and where the vacuum energy comes from which leads to the accelerated expansion. At last, we also applied our inhomogeneous model to find the tunneling amplitude of the universe from nothing. We found  the spatial fluctuations caused by the inhomogeneous vacuum lead to faster tunneling while the tunneling amplitude is dependent on the spatial locations.

There are still some issues that we have not addressed with our model in this paper. First, in our model, we take the simplest potential for the scalar field. However, our method which describes the inhomogeneity by modifying Green's function is general. Therefore, an improvement for our model is to consider more complicated case, such as grand unified theory. Second, no higher order correlations, such as one-loop correlations, are considered here in our model. Thus, the loop correction should be involved in our model and then more accurate effective potential can be obtained. Third, it would interesting to consider the cosmological consequences of the new inflation scenario suggested here especially the density fluctuations as the seed for the large scale structure formation and the related fluctuation power spectrum. These are closely associated to the current observations of the microwave background radiation and large scale density map.

\section*{acknowledgements}
Jin Wang would like to thank NSF Phys. 76066 for partial support

\bibliographystyle{unsrt}
\bibliography{mybib}

\begin{thebibliography}{10}
\bibitem{Notea}
In fact, here, $\phi ^4$ should be replaced by 4-point correlation function,
  which includes at least one four-point interaction. However, once $O(\lambda
  )$ and higher order diagrams are omitted, ${G_R}^2$ is still good enough to
  be used for the approximation.
\bibitem{Noteb}
Strictly speaking, the imaginary time method should be applied to discuss the
  modification of $A$ and $B$, but it makes no difference to our result. For
  details of the imaginary time, see Ref.\cite {ashok}.
\end{thebibliography}

\end{document}